\documentclass[twocolumn]{revtex4-1}

\usepackage{amsmath}
\usepackage{amsfonts}
\usepackage{graphicx}

\begin{document}

\title{An Empty Chiral Rotation for the Adler-Bell-Jackiw Anomaly}

\author{Israel Weimin Sun}
\email{sunwm@nju.edu.cn}

\affiliation{School of Physics, Nanjing University, Nanjing~210093, China}

\date{\today}

\begin{abstract}
This is an article which intends to shake down the traditional belief that the celebrated Adler-Bell-Jackiw anomaly stems from the chiral rotation non-invariance 
of the fermionic measure. The fermionic functional integration measure in quantum field theory should be defined so as to reproduce the standard Feynman diagrammatic 
expansion. This implies that a plain definition of the fermionic measure automatically serves such a purpose. 
A dilemma then arises: how could one identify the ABJ anomaly as a nontrivial Jacobian factor for a chiral transformation ?  The true answer is indeed surprising 
and unexpected, that is, the Jacobian factor is actually a random and indeterminate object, hence it carries no physical information.
A true explanation for the ABJ anomaly is suggested.
\end{abstract}

\maketitle

{\it Introduction.}
The quantum anomaly is a physical phenomenon deeply rooted in relativistic quantum field theories, whose appearence is manifested as the breakdown of some classically valid symmetry after the field quantization procedure. The so-called Adler-Bell-Jackiw anomaly (or the chiral anomaly) \cite{Book-Adler-Jackiw}  
in the abelian quantum electrodynamics is just an example in this category.  In the framework of the renowned path-integral quantization the {\it raison d'etre} of the ABJ anomaly 
is ascribed to the chiral rotational non-invariance of the fermionic functional integration measure. This seemingly splendid success was established by 
Kazuo Fujikawa in his seminal work in 1979 \cite{Fujikawa-PRL}. A common wisdom of the physics community is that Fujikawa's 
theoretical framework is a universal one since it has covered all the quantum anomalies known in field theory \cite{Fujikawa-book}.

But unfortunately there seems to be a huge stone in the way of our understanding of all these physical matters. It is based on a rather simple (but fatal) observation: in our present path-integral formalism of spinor QED (and any other field theoretical models of such types), by means of separating the total classical field action into a free-field part plus an interacting part: $S[{\bar \psi},\psi,A_\mu]=S_0[{\bar \psi},\psi,A_\mu]+S_{int}[{\bar \psi},\psi,A_\mu]$,
one could adopt
a simplest definition of the Grassmann fermionic functional integration measure
(such as that in a free fermion theory), together with a gauge-fixed path-integral measure of the $U(1)$ gauge field, to establish a standard set of Feynman rules which gives
an order-by-order perturbative expansion of all the $n$-point Green functions for all types of elementary interpolating field insertions (i.e., excluding those of composite field insertions), which after some appropriate renormalization procedure would yield a complete set of transition amplitudes of all QED processes, i.e., all the S-matrix elements,
within some physically reasonable accuracies. In short, a plain definition of the Grassmann fermionic functional integration measure, such as the one in a non-interacting situation, plus some standard definition of the path-integral measure for the gauge-field sector, would yield the standard machinery of perturbative QED theory through our
standard practice in theoretical physics, even though it strongly disagrees with the equally "standard" choice of defining the Grassmann fermionic fields (and the relevant
fermionic measure) in one's computation of the ABJ anomaly in the framework of path-integral method, whose basic spirit is to produce something representing the ABJ anomaly by
"rotating out" an $A_\mu$-dependent functional Jacobian factor for a chiral rotation of the classical fermion field !

In fact, such a disagreement is not just a purely formal difference, since the standard mathematical practice tells us an absolute truth: a plain definition of the classical fermion fields $({\bar \psi},\psi)$, which has nothing to do with the gauge-field sector, could merely yield an $A_\mu$-independent Jacobian factor for a finite chiral rotation so that
one could no longer expect to see an "ABJ anomaly" term by means of such a calculation ! However, an entanglement with the gauge-field sector is actually the
 "absolute {\it raison d'etre}" for Fujikawa's evaluation of the ABJ anomaly using the path-integral idea.
So, such a dilemma seems to be a severe damaging point against the standard path-integral explanation of the physical origin of the ABJ anomaly which has been firmly established for
40 years !

Needless to say, the ABJ anomaly is definitely there, and all these apparent difficulties need to have a true explanation. In this short note, I intend to provide a full answer to all
these issues. I shall show that: (1) a natural definition of the $({\bar \psi},\psi)$ field in spinor QED (actually in all other field theoretical models) should not be entangled
with the existence of the $U(1)$ gauge field degrees of freedom, although Fujikawa's original
"gauge-field-sector-entangled" definition could also be used to define everything; (2) an evaluation of Fujikawa's original "functional Jacobian factor" within a larger family
of gauge-invariant regularizations reveals that this Jacobian is actually a totally random and indeterminate expression, i.e, an object which is too random to represent any physical reality; (3) as a result of all these facts, one should forget about the traditional Old Dogma that the so-called "chiral rotation
non-invariance" of the fermionic measure is the {\it raison d'etre} for the celebrated ABJ-anomaly in field theory. Finally, I would like to
point out that a natural explanation of the ABJ anomaly in the path-integral framework does exist, in which the so-called "chiral rotation non-invariance" issues are
simply avoided, so that everything is saved from being damaged.

{\it An analysis of the real situation: how should one define a fermionic functional measure?  }
As usual I will consider a spinor QED theory consisting of one species of Dirac fermion and an abelian gauge field with which it has a nontrivial
interaction. In all applications of the path-integral method, one needs to input a primary definition of the fermionic measure into the whole formalism.
In fact, the usual perturbative expansion of all $n$-point functions of QED is solely based on the Wick's theorem, namely, the expression for the many-point moments of a free Gaussian measure. For the Dirac fermion sector, the relevant Grassmann integration with sources has the form
\begin{equation}\label{Gaussian-type-integration}
\int\mathcal{D}{\bar \psi}\mathcal{D}\psi e^{-{\bar \psi}K_0 \psi + {\bar \psi}\eta+{\bar \eta}\psi}.
\end{equation}
Practically, this Gaussian integration would generate everything one needs in a nontrivial perturbative calculation. No matter how one chooses to define such a symbolic expression,
it should posses two basic properties:~(1) translation-invariance of the form $\int\mathcal{D}{\bar \psi}\mathcal{D}\psi F[{\bar \psi},\psi]=\int\mathcal{D}{\bar \psi}\mathcal{D}\psi F[{\bar \psi}+{\bar \psi}_0,\psi+\psi_0]$;~(2) under a linear transformation of $({\bar \psi},\psi)$, the functional Jacobian equals the inverse of the relevant determinant factor.

A practical definition of such a Grassmann integration consists in introducing a countable set of Grassmann generators and expanding the $({\bar \psi},\psi)$ field as follows
\begin{eqnarray}\label{natural-fermion-field-definition}
\nonumber   \psi(x)&=& \sum_{n} \xi_n(x) a_n  \\
 {\bar \psi}(x)&=& \sum_{n}  {\bar b}_n \xi^\dagger_n(x)
\end{eqnarray}
where $\{\xi_n(x)\}$ is some set of basis functions. One then tacitly assumes
$\mathcal{D}{\bar \psi}\mathcal{D}\psi = \prod_n d {\bar b}_n d a_n$. This definition guarantees the above mentioned two properties.

All we need to know is just the fundamental Gaussian integration identity:
\begin{equation}\label{Gaussian-integration-identity}
\int\mathcal{D}{\bar \psi}\mathcal{D}\psi e^{-{\bar \psi}K_0 \psi + {\bar \psi}\eta+{\bar \eta}\psi}=\int\mathcal{D}{\bar \psi}\mathcal{D}\psi e^{-{\bar \psi}K_0 \psi}\times
e^{{\bar \eta}K^{-1}_0\eta},
\end{equation}
which is readily established by a simple translation of variables. Here a most essential fact appears:  (\ref{Gaussian-integration-identity}) is a universal formula, i.e.,
it holds irrespective of the concrete choice of
the basis function in (\ref{natural-fermion-field-definition}). Then, an expansion of both sides of (\ref{Gaussian-integration-identity}) in terms of the anticommuting sources yields the usual Wick's theorem of the following format
\begin{eqnarray}\label{Wick}
\nonumber &&\frac{\int \mathcal{D}{\bar \psi}\mathcal{D}\psi \psi(x_1){\bar \psi}(x_2)\cdots \psi(x_{2N-1}){\bar \psi}(x_{2N})e^{-S_0[{\bar \psi},\psi]}}{\int  \mathcal{D}{\bar \psi}\mathcal{D}\psi e^{-S_0[{\bar \psi},\psi]}} \\
&=& \sum_{{\rm pairs}}\prod S_F(x_i-x_j).
\end{eqnarray}
This expression is valid for ${\it all}$ choices of the basis function $\{\xi_n(x)\}$.

Now let me turn to Fujikawa's calculation of the ABJ anomaly. When a classical background gauge field $A_\mu$ exists, Fujikawa insists that one should expand the classical fermion fields in a gauge-field-sector-entangled manner
\begin{eqnarray}\label{Fujikawa-fermion-field-definition}
\nonumber   \psi(x)&=& \sum_{n} \varphi_n(A;x)a_n  \\
 {\bar \psi}(x)&=& \sum_{n}  {\bar b}_n \varphi_n^\dagger(A;x),
\end{eqnarray}
where his basis functions $\varphi_n(A;x)$ are just the complete set of eigenfunctions of the Euclidean Dirac operator: $i\not\! D (A)\varphi_n(A;x)=\lambda_n(A)\varphi_n(A;x)$.
The merit of the expansion (\ref{Fujikawa-fermion-field-definition}) is merely a formal mathematical beauty: the gauge-invariant fermionic action has an explicitly diagonalized form so that a direct application of the Berezin integration rule yields
\begin{eqnarray}\label{Functional-DET}
\nonumber &&\int\mathcal{D}{\bar \psi}\mathcal{D}\psi e^{-\int d^4 x {\bar \psi}(\not\! D (A)+m)\psi} \\
\nonumber &=&\int \prod_n d{\bar b}_n da_n e^{-\sum_n (-i\lambda_n(A) +m){\bar b}_n a_n} \\
&=& \det(\not\! D (A)+m).
\end{eqnarray}
However, I intend to point out a simple mathematical fact: any definition of the form (\ref{natural-fermion-field-definition}) with an ${\it arbitrarily}$ chosen basis function $\{\xi_n(x)\}$ is equally capable of producing the same functional determinant in (\ref{Functional-DET})! The argument is simple. One could formally separate out the $j\cdot A$
interaction term from the gauge-invariant fermionic action and obtain
\begin{eqnarray}\label{many-steps}
\nonumber &&\mathcal{N}\int\mathcal{D}{\bar \psi}\mathcal{D}\psi  e^{-\int d^4 x {\bar \psi}(\not\! D (A)+m)\psi} \\
\nonumber &=& \mathcal{N}\int\mathcal{D}{\bar \psi}\mathcal{D}\psi  e^{-\int d^4 x {\bar \psi}(\not\! \partial +m)\psi} \sum_n \frac{(-1)^n}{n!}
\big[ \int (ie{\bar \psi}\gamma_\mu \psi A_\mu)\big]^n \\
&&
\end{eqnarray}
A formal functional integration using the Wick's theorem (\ref{Wick}) just produces a formal Feynman diagram expansion which represents exactly the functional determinant in (\ref{Functional-DET}).
Therefore, it is clear that every choice of the basis function in (\ref{natural-fermion-field-definition}) can do the job, while Fujikawa's definition (\ref{Fujikawa-fermion-field-definition}) is just a specific one among all these possible choices.

In the same manner, one could also readily show that the whole machinery of perturbative QED theory, i.e., the relativistic covariant Feynman diagrammatic
expansion of all its $n$-point Green functions (or the Schwinger functions in a Euclidean spacetime) could be established on the bases of the definition (\ref{natural-fermion-field-definition}) and the appropriate definition of the functional integration measure of the gauge-field sector. The most important point is that all choices of the basis function $\{\xi_n(x)\}$ in (\ref{natural-fermion-field-definition}) are equally capable of achieving this purpose.

Then, since a natural definition of the fermionic measure suffices and the relevant functional Jacobian for a chiral transformation is necessarily $A_\mu$-independent, how could one "explain" the very existence of the ABJ anomaly within the framework of the path-integral formalism ?
If one accepts the basic fact that the choice of the basis
function in (\ref{natural-fermion-field-definition}) is arbitrary, one needs to make a universal judgment: whether Fujikawa's original evaluation of the functional Jacobian factor
is absolutely correct so that his background-gauge-field-dependent definition (\ref{Fujikawa-fermion-field-definition}) should be a mandatory choice, or
one needs to invoke some other mechanism to explain the existence of the ABJ anomaly ?

Needless to say, such a judgment should not be self-contradictory or misleading in any sense.
In the following I will provide a careful and critical analysis of Fujikawa's original evaluation of the Jacobian factor and show that his regularized "Jacobian factor" is a totally ambiguous object and thus should not carry any physical information with
it. I hope my argument is clear enough to wash away all the misunderstanding of this issue for so many years.

{\it The evaluation of the Jacobian factor: how to make a divergent series to be a convergent one ?}
Here let me first recall some basic facts. If one considers a chiral transformation of $({\bar \psi},\psi)$
\begin{eqnarray}
\nonumber \psi'(x)&=& e^{i \alpha(x)\gamma_5}\psi(x) \\
{\bar \psi}'(x)&=&{\bar \psi}(x)e^{i \alpha(x)\gamma_5},
\end{eqnarray}
the relevant Jacobian factor ${\it \grave{a}~la}$ Fujikawa would be
\begin{equation}\label{Fujikawa-Jacobian}
J= e^{-2i \int d^4x \alpha(x)\sum_n \varphi^\dagger_n(A;x)\gamma_5 \varphi_{n}(A;x)}.
\end{equation}
When one uses a natural definition (\ref{natural-fermion-field-definition}), the corresponding Jacobian factor is obtained by a simple substitution $\varphi_{n}(A;x)\rightarrow \xi_n(x)$ in (\ref{Fujikawa-Jacobian}).

As it stands, the formal summation
\begin{equation}\label{formal-summation}
S[A;x]=\sum_{n} \varphi^\dagger_n(A;x)\gamma_5 \varphi_{n}(A;x)
\end{equation}
is an ill-defined process. In order to extract something meaningful from it, one needs to regularize it. In a formal sense, the summation (\ref{formal-summation}) is a gauge-invariant
process, since $\varphi_n(A^U)= U \varphi_n(A)$, and a good regularization should respect the gauge-invariance. Fujikawa chooses the following regularization \begin{equation}\label{Fujikawa-regularization}
S_{reg}[A]= \lim_{M^2 \rightarrow \infty}\sum_{n} \varphi^\dagger_n(A;x)\gamma_5 \varphi_{n}(A;x)e^{-\lambda^2_n(A)/M^2},
\end{equation}
where the large $\lambda_n$ contributions in the summation are suppressed by the Gaussian cut-off factor. This regularization is gauge-invariant because of
$\lambda_n(A^U)=\lambda_n(A)$. The regularized sum is calculated as
\begin{equation}\label{ABJ}
S_{reg}[A]=\frac{e^2}{32 \pi^2}\epsilon_{\mu\nu\rho\sigma} F_{\mu\nu}F_{\rho\sigma},
\end{equation}
which produces the expected ABJ anomaly.

This is the standard story. However, there is no reason to believe that the regularization (\ref{Fujikawa-regularization}) should be the unique one. In fact, all one needs to do is to speed up the rate of convergence for the series (\ref{formal-summation}) by damping its large eigenvalue contributions. If one works in the framework of gauge-invariant regularizations, one could try to regularize it in a different manner
\begin{equation}\label{My-regularization}
S_{reg|(A\rightarrow {\tilde A})}[A]= \lim_{M^2 \rightarrow \infty}\sum_{n} \varphi^\dagger_n(A;x)\gamma_5 \varphi_{n}(A;x)e^{-\lambda^2_n({\tilde A})/M^2},
\end{equation}
where $\lambda_n(\tilde{A)}$ is the relevant eigenvalues of some different Dirac operator $i\not\! D ({\tilde A})$. Here one should assume that the two set of eigenvalues $\{\lambda_n(A)\}$ and $\{\lambda_n(\tilde{A)}\}$ are in one-to-one correspondence with each other and at the same time both $\lambda_n(A)$ and $\lambda_n(\tilde{A)}$ grow unboundedly when $n$
gets large. In order to guarantee this property, one needs to assume the pair of gauge field configurations $(A,\tilde{A})$ to be sufficiently close to each other.
Needless to say, such a deformation $A \rightarrow {\tilde A}$ is not a pure gauge transformation, i.e., it
should change the field strength $F_{\mu\nu}$. I will now show that this method could serve as a gauge-invariant regularization when structured appropriately.

The recipe is very simple.  One first notes that the eigenvalue $\{\lambda_n(A)\}$ only depends on the "gauge orbit" (i.e., the gauge equivalence class) to which a particular gauge field configuration $A$ belongs, so that one could effectively write $\lambda_n(A) = \lambda_n([A])$ where $[A]$ denotes the corresponding "gauge equivalence class". With this at hand, one then arbitrarily picks a map $F$ from the gauge orbit space to itself
whose effect is to establish a "physical correspondence" between the various gauge orbits. So one writes effectively $F: [A] \mapsto [{\tilde A}]=[A]^F$. Here one has to make the additional assumption that under the $F$-action
each gauge orbit $[A]$ is only slightly changed so that the two eigenvalue sets $\{\lambda_n(A)\}$ and $\{\lambda_n(\tilde{A)}\}$
are appropriately matched. Therefore, one sees clearly a chain of correspondence: $A \mapsto [A] \mapsto [A]^F=[{\tilde A}]$, thus each one of the
eigenvalues $\lambda_n({\tilde A})=\lambda_n([{\tilde A}])=\lambda_n([A]^F)$ could be effectively regarded as a "gauge-invariant body"  made up of the background gauge field $A$.

With all such preparations, I can generalize the original Fujikawa's regularization (\ref{Fujikawa-regularization}) to a whole family of new regularizations which I call $F$-action
Gaussian cutoff regularization, or $F$-regularization for short. When the gauge-orbit-space map $F$ is degenerate to the identity map, this regularization would coincide with
Fujikawa's original one, hence it is an essential enlargement of the original framework. Such an $F$-regularization, as I mentioned previously, preserves explicit gauge invariance,
and all of my further analysis will be based on it.

Now let me describe all the necessary steps. First of all, since $A$ and $\tilde{A}$ are sufficiently close to each other, one naturally assumes the two sets of basis functions, namely, $\{\varphi_n(A;x)\}$ and $\{\varphi_n(\tilde{A};x)\}$ are exactly in one-to-one correspondence. They are connected by a formal unitary transformation which reads
\begin{equation}\label{unitary-transformation}
\varphi_n(A;x)=\sum_{m}K_{nm}\varphi_m(\tilde{A};x).
\end{equation}
Then, using the relation (\ref{unitary-transformation}) I establish a chain of derivations
\begin{eqnarray}\label{natural}
\nonumber && \sum_n \varphi^\dagger_n(A;x) \gamma_5\varphi_n(A;x)e^{-\lambda^2_n({\tilde A})/M^2} \\
\nonumber &=&\sum_n \big( \sum_m  \varphi^\dagger_m(\tilde{A};x)K^{*}_{nm} \big)\gamma_5 \big( \sum_{{\bar m}} K_{n{\bar m}}\varphi_{{\bar m}}(\tilde{A};x)  \big)e^{-\lambda^2_n({\tilde A})/M^2} \\
\nonumber &=& \sum_{m~{\bar m}} \varphi^\dagger_m(\tilde{A};x)\gamma_5 \varphi_{{\bar m}}(\tilde{A};x) \big(\sum_n K^{*}_{nm}K_{n{\bar m}}e^{-\lambda^2_n({\tilde A})/M^2} \big) \\
\nonumber &=& \sum_{m~{\bar m}} \varphi^\dagger_m(\tilde{A};x)\gamma_5 \varphi_{{\bar m}}(\tilde{A};x)\big(\delta_{m{\bar m}}e^{-\lambda^2_m({\tilde A})/M^2}+ R_{m{\bar m}} (M^2)\big) \\
&=& \sum_{m} \varphi^\dagger_m(\tilde{A};x)\gamma_5 \varphi_{m}(\tilde{A};x)e^{-\lambda^2_m({\tilde A})/M^2}+ {\rm Remainder}.
\end{eqnarray}
Finally, I shall argue that $Remainder(M^2)\stackrel{M^2 \rightarrow \infty}{--\longrightarrow} 0 $ so that $S_{reg|(A\rightarrow {\tilde A})}[A]= S_{reg}[A]|_{A\rightarrow {\tilde A}}$.

To see why this is so, let me rewrite $R_{m{\bar m}}$ as
\begin{equation}\label{R-part}
R_{m{\bar m}}(M^2)=\sum_n \big(K^{*}_{nm}K_{n{\bar m}}-\delta_{nm}\delta_{n{\bar m}}\big)e^{-\lambda^2_n({\tilde A})/M^2}.
\end{equation}
I will show that $\lim_{M^2 \rightarrow \infty}R_{m{\bar m}}(M^2)=0$. The argument is like this. Denote $a_n= K^{*}_{nm}K_{n{\bar m}}-\delta_{nm}\delta_{n{\bar m}}$.
Since the $K$ matrix is a unitary one, one has automatically $\sum_n a_n=0$. Now, if the sequence $\{a_n\}$ is absolutely summable, i.e., $\sum_n |a_n| < \infty$, then I could invoke
the so-called dominated convergence theorem to interchange the order of the summation and the $M^2 \rightarrow \infty$ limit:
\begin{equation}\label{limit}
\lim_{M^2 \rightarrow \infty} \sum_n a_n e^{-\lambda^2_n({\tilde A})/M^2}=\sum_n  a_n =0,
\end{equation}
which is the desired result. This absolute summability can be readily verified. For $m={\bar m}$, since $\sum_n |K_{nm}|^2=1$, the sequence $\{K^{*}_{nm}K_{nm}\}$ is absolutely summable, hence this is also the case for $\{ K^{*}_{nm}K_{nm}-\delta_{nm}\delta_{nm}\}$.
For $m\not={\bar m}$, the absolute summability of $\{K^{*}_{nm}K_{n{\bar m}}\}$ can be established in the following way. First note that both of the two sequences $\{ |K^{*}_{nm}| \}$ and $\{ |K_{n{\bar m}}| \}$ are square summable, i.e., they all belong to $l^2$. Then, an elementary application of the Cauchy-Schwarz inequality on the $l^2$ space gives
\begin{equation}\label{inequality}
\sum_n |K^{*}_{nm}K_{n{\bar m}}|\leq \big(\sum_n |K^{*}_{nm}|^2 \big)^{1/2}\big(\sum_n |K_{n{\bar m}}|^2 \big)^{1/2}=1,
\end{equation}
which establishes the desired absolute summability. With the result $\lim_{M^2 \rightarrow \infty}R_{m{\bar m}}(M^2)=0$ at hand, one could safely conclude that the "Remainder"
term in (\ref{natural}) should vanish in the $M^2 \rightarrow \infty$ limit.

Now, since to a large extent the physical correspondence $F: [A] \mapsto [{\tilde A}]$ is arbitrary, one may conclude that
the regularized "Jacobian factor" is a totally random and indeterminate  expression, i.e, an object which is too random to represent any physical reality. This implies a crucial fact, that is, Fujikawa's original evaluation of the functional Jacobian is a fake process, hence one is unable to extract anything sensible from this process which could stand for a genuine ABJ anomaly term. Of course, nobody is willing to say it is
a real damaging point against the existing knowledge of quantum anomalies, since the ABJ anomaly is a true physical existence and should never be washed away in the realm of physics.
Nevertheless, an essential modification to our conventional wisdom of QFT seems to be unavoidable, that is, one should abandon the old belief that the ABJ anomaly is caused by the so-called "chiral rotation non-invariance" of the fermionic functional integration measure, since the relevant "Jacobian factor" is totally random, even though it has a perfectly finite expression ! This opinion should not be rejected by the long-period practice of high energy physics experiments. Any endeavor in theoretical high energy
physics should have its foundation in genuine mathematics, instead of just living on purely technical advances.

With all the previous facts being established, a last question necessarily arises: how could one identify a true mechanism in the path-integral formalism which correctly explains
the physical existence of the celebrated ABJ anomaly? Such a mechanism does exist, which is clearly described in the monograph \cite{Fujikawa-book}. One only needs to introduce
a compensating Pauli-Villars regulator field with a large mass to regularize everything. In this method, the ABJ anomaly term emerges naturally as the result of the compensating
effect of
the two types of loop diagrams: the one corresponding to the original fermion field together with an opposite one corresponding to the regulator field.
Now, since the regulator field should be a bosonic one, the total "Jacobian factor" equals one automatically, thus in this scheme one needs not to resort to a nontrivial Jacobian
to explain the physical existence of the ABJ anomaly.
Therefore, the ABJ anomaly is definitely there, but the conventional path-integral interpretation for its origin needs to be changed. In connection with this, I would also like
to mention a recent e-print \cite{Sun} in which it is shown that for the so-called transverse anomalies
Fujikawa's path-integral method would yield a result different from that of the one-loop perturbative calculation. This also shows the weakness of Fujikawa's approach.

{\it Concluding remarks.}
In this note I show that the definition of fermionic functional integration measure in quantum field theory should not be entangled with the gauge-field sector.
It is also rather unexpected to see that Fujikawa's path-integral evaluation of the ABJ anomaly is actually a fake process since his "Jacobian factor" is too random to
represent any physical reality. I believe this trend is a healthy one even though it has shaken the old basis of one's understanding of quantum anomalies established
in the past 40 years. The more general situations seem to be essentially the same. I plan to discuss them in the future works.

I thank Fan Wang for useful conversations.
This work is supported in part by the Natural Science Funds of Jiangsu Province of China under Grant No. BK20151376.


\begin{thebibliography}{99}
\bibitem{Book-Adler-Jackiw} For a splendid account of the historical matters of quantum anomalies in field theory, see the two contributed articles of Stephen L. Adler
together with Roman Jackiw in {\it 50 Years of Yang-Mills Theory}, edited by Gerardus 't Hooft, (World Scientific, 2005).
\bibitem{Fujikawa-PRL} K. Fujikawa, Phys. Rev. Lett. {\bf 42}, 1195 (1979).
\bibitem{Fujikawa-book}Kazuo Fujikawa and Hiroshi Suzuki, {\it Path Integrals and Quantum Anomalies} (Clarendon Press, Oxford 2013).
\bibitem{Sun} Israel Weimin Sun, arXiv: 1904.04638 [physics.gen-ph].
\end{thebibliography}
\end{document}